\documentclass[showpacs,amsmath,amssymb,aps,prb,twocolumn]{revtex4-1}

\usepackage{graphicx}
\usepackage{dcolumn}
\usepackage{bm}
\usepackage{color}

\begin{document}

\title{Vibrational effects in x-ray absorption (XAS) and resonant inelastic x-ray scattering (RIXS) using a semiclassical scheme}

\author{Mathias P. Ljungberg}

\affiliation{Donostia International Physics Center, Paseo Manuel de Lardizabal, 4. E-20018 Donostia-San Sebasti\'{a}n, Spain}

\begin{abstract}
A new method is presented for describing vibrational effects in x-ray absorption spectroscopy (XAS)  and resonant inelastic x-ray scattering (RIXS) using a combination of  the classical Franck-Condon (FC) approximation and classical trajectories run on the core-excited state. The formulation of RIXS is an extension of the semiclassical Kramers-Heisenberg (SCKH) formalism of Ref. \onlinecite{Ljungberg_2010} 
 to the resonant case, retaining approximately the same computational cost. 
To overcome difficulties with connecting the absorption and emission processes in RIXS the classical FC approximation is used for the absorption, which is seen to work well provided that a zero-point-energy correction is included. 
 In the case of core-excited states with dissociative character the method is capable of closely reproducing the main features for one-dimensional test systems, compared to the quantum mechanical formulation.
Due to the good accuracy combined with the relatively low computational cost, the method has large potential of being used for complex systems with many degrees of freedom, such as  liquids and surface adsorbates.
\end{abstract}

\maketitle

\section{Introduction}

X-ray absorption spectroscopy (XAS) and and x-ray emission spectroscopy (XES) are important tools for investigating the electronic structure of molecules, liquids, solids and surface adsorbates \cite{Stohr_book, Nilsson_2004, Gelmukhanov_1999, Ankudinov_1998}. 
Due to the involvement of a core level, these spectroscopes are local and element specific and can give valuable information about local molecular geometries and chemical bonding. The two spectroscopies are complementary since XAS probes the unoccupied states and XES the occupied states, projected on the targeted species. Resonant inelastic x-ray scattering (RIXS), also called resonant XES,  combines the absorption and emission in a single process and can give additional information about the symmetry of states and a more sensitive control of the emission by detuning the incoming radiation with respect to the resonances.  
 For systems where the core excited (or core ionized) state is dissociative, or has a very distorted potential energy minimum,  large vibrational effects occur upon the creation of the core hole. In XAS this manifests by the Franck-Condon profile --- the series of peaks coming from excitation to higher vibrational states in the final electronic state --- merging to a broad distribution mapping the ground state vibrational wave function \cite{Gelmukhanov_1999}. In XES the effects are more profound as the emission can take place on severely distorted geometries.
 The proper way to describe these effects are as vibrational interference in the eigenstate picture, or as wave packet dynamics in the time domain, both of which can be described by the Kramers-Heisenberg (KH) formula in different representations \cite{Gelmukhanov_1999}.  
A fully quantum mechanical description of the vibrational effects is only feasible for a few degrees of freedom and the potential energy surfaces (PESes) must be pre-computed, which makes realistic models of XES in, for example, liquids very hard. For this reason, semiclassical models that use ensembles of classical trajectories to approximately capture the effects of dissociation have been devised, and used to compute the XES of for example liquid water \cite{Tokushima_2008, Odelius_2009, Odelius_2009_2, Ljungberg_2010}. These methods are not rigorous in that there is no exact quantum limit and the approximations are not well-defined \cite{Mukamel_1982}. However, if the relevant effects of bond distortions or dissociation on the spectra are reproduced, such methods can be very valuable in the interpretation of experimental result for complicated systems such as liquids, since 
an arbitrary
number of vibrational degrees of freedom
can be treated. A semiclassical approximation of the KH formula, denoted the semiclassical Kramers-Heisenberg (SCKH) method \cite{Ljungberg_2010}, was devised  for non-resonant  XES, that is for the case where the system is core ionized with the extra electron removed from the system, and was shown to closely reproduce the quantum mechanical spectrum for a one-dimensional model of a water dimer. Recently, the method was used to compute XES of liquid methanol  \cite{Ljungberg_2017}, and ethanol \cite{Takahashi_preprint_2017}, using realistic clusters containing seventeen molecules
that were 
extracted 
from molecular dynamics simulations. 
The  comparison to experiment was seen to be good, including the isotope effect (for methanol) that was seen to be of purely dynamical origin.  
In the present publication, the SCKH method is extended to describe RIXS, which also makes a further analysis of vibrational effects in XAS necessary. Since the main objective is to describe liquids that have large vibrational effects due to (locally) dissociative core-excited states the approximations used will be tailored for this purpose. For XAS, the classical Franck-Condon (FC) approach is seen to work quite well, 
provided that
the zero-point energy in the initial state is included.
For RIXS, a combination of the classical Franck-Condon approximation with excited state dynamics
is seen to give very good results compared to the quantum mechanical description for one-dimensional test cases. 
The new method has a computational cost comparable to the non-resonant SCKH method and is seen to have similar accuracy, which has the potential to open the door to accurate modeling of RIXS for complex systems, such as liquids, that display large vibrational effects.  
 
\section{Theory}

The absorption and emission processes occurring in RIXS will be treated in a one-step manner through the Kramers-Heisenberg formula in order to include the vibrational effects that come from the interference of intermediate vibrational states \cite{Gelmukhanov_1977, Gelmukhanov_1999, Ljungberg_2011}. From the quantum mechanical description a semiclassical approximation will be developed, which is similar to that in Ref. \onlinecite{Ljungberg_2010} but extended to the resonant case. 

As it is not trivial to approximate a quantum mechanical expression to one where classical dynamical trajectories can be used, a series of approximations must be made, and it can not be expected that all details can be faithfully reproduced. Nevertheless a set of properties that are desirable for such an approximation are 

\begin{enumerate}
\item  The formalism must be able to describe the vibrational effects for dissociative intermediate states in XES. 
\item The integrated cross section with respect to either the emission energy (proportional to XAS) or the incoming energy (leading to broadband excited XES) must be reproduced.
\item Extra spurious features that can obscure interpretation must be avoided.
\item In absence of vibrational effects, the method should reduce to the normal Kramers-Heisenberg formula for electronic states.
\item In terms of sampling of initial conditions, fast convergence is desired. For large systems the number of initial conditions will have to be moderate in order to minimize the computational cost.
\end{enumerate}
   
Point number two above requires a good description of the vibrational effects in the x-ray absorption (XAS) along with the non-resonant XES (that is,  ionized, or resonant with broadband excitation). Furthermore these two have to be connected in a consistent manner. As the semiclassical non-resonant method has been shown to work very well for test systems \cite{Ljungberg_2010, Ljungberg_2017}, and also for a realistic model of liquid methanol  \cite{Ljungberg_2017} and ethanol \cite{Takahashi_preprint_2017}, it  will be retained as far as possible in the resonant formulation. In the following sections the quantum mechanical cross section for XAS and RIXS are first rewritten in the time domain, which will be the starting point for the semiclassical approximations that will later be developed. 

\subsection{Time-domain formulation of XAS}

The XAS 
cross section at incoming frequency $\omega$ is given by Fermi's golden rule
\begin{equation}
\sigma(\omega) \propto \sum_{f} \left | \langle i | D | f \rangle  \right |^2 \delta(\omega-E_{fi}) \, ,
\label{eq:XAS_cross_section}
\end{equation}
where $| i \rangle$ and  $| f \rangle$ are the initial and final states (including both electronic and nuclear degrees of freedom) that are eigenstates of the full Hamiltonian, $H$,  and $E_{fi} = E_f-E_i$ is the energy difference between the states. The transition operator $D$ can 
be written as $D=d$ (or $D =  [d, H ]$ depending which light-matter coupling is used), with $d = \mathbf{e} \cdot \mathbf r$ that depends on the electric field $\mathbf{e}$ and the position operator $\mathbf r$. Atomic units are used throughout.
The first step in transforming Eq. (\ref{eq:XAS_cross_section}) is to put 
the delta function inside the square using the representation
\begin{equation}
\delta(\omega) = \lim_{\Gamma \to 0} \frac{\Gamma_f}{\pi} \left | \frac{1}{\omega + i\Gamma_f} \right |^2 \, , 
\label{eq:delta_function}
\end{equation}
which for a small, but finite, value of the broadening parameter $\Gamma_f$ leads to
\begin{equation}
\sigma(\omega) \propto \frac{\Gamma_f}{\pi}  \sum_{f} \left | \frac{ \langle i | D | f \rangle}{\omega - E_{fi}+ i\Gamma_f} \right |^2 \, .
\label{eq:XAS_sigma_square}
\end{equation}
 It is then possible to switch from frequency to time domain by using the identity
\begin{equation}
\frac{1}{\omega + i\Gamma_f}  =-i \int^{\infty}_{0} dt e^{i(\omega +i\Gamma_f) t} \, ,
\label{eq:resolvent}
\end{equation}
which transforms the expression inside the square in Eq. (\ref{eq:XAS_sigma_square}) to
\begin{align}
  \frac{ \langle i | D | f \rangle}{\omega - E_{fi}+ i\Gamma_f} &=  -i \int^{\infty}_{0} dt \langle i | e^{i E_i t} D e^{-i E_f t} | f \rangle  e^{i(\omega +i\Gamma) t} \\ 
 &=  -i \langle i | \int^{\infty}_{0} dt  e^{i H t} D e^{-i H t}  e^{i(\omega +i\Gamma_f) t}  | f \rangle
 \label{eq:XAS_time_2} \\
 &=  \langle i | D(\omega) | f \rangle \, .
 \label{eq:XAS_time_3}
\end{align}
In Eq. (\ref{eq:XAS_time_2}) the identities $\langle i | e^{iE_i t } = \langle i | e^{iH t }$ and  $e^{-iE_f t } | f \rangle = e^{-iH t } | f \rangle$ were used. The half-Fourier transformed dipole operator in Eq. (\ref{eq:XAS_time_3}) can be written 
\begin{equation}
D(\omega) =  -i\int^{\infty}_{0} dt  D(t) e^{i(\omega +i\Gamma_f) t} \, ,
\label{eq:D_omega_FT}
\end{equation}
using the Heisenberg representation of a time-dependent operator $D(t) = e^{i H t} D(0) e^{-i H t}$. Since the state energies no longer appear in this formula the square in the cross section in Eq. (\ref{eq:XAS_sigma_square}) can be expanded and the sum over $| f \rangle$ states removed by using the resolution of the identity $\sum_f | f\rangle \langle f | = 1$, yielding
\begin{align}
\sigma(\omega) &\propto \frac{\Gamma_f}{\pi}  \sum_{f} \left | \langle i | D(\omega) | f \rangle \right |^2 \\
&= \frac{\Gamma_f}{\pi}  \langle i | D(\omega)  \sum_{f} | f \rangle \langle f | D^{\dagger}(\omega) | i \rangle \\
&= \frac{\Gamma_f}{\pi}  \langle i | D(\omega) D^{\dagger}(\omega) | i \rangle   \, .
\label{eq:XAS_sigma_eigfree_3}
\end{align}
The ground state expectation value in Eq. (\ref{eq:XAS_sigma_eigfree_3}) can be generalized to a thermal average as
\begin{equation}
\sigma(\omega) \propto \frac{\Gamma_f}{\pi}  \left \langle  D(\omega) D^{\dagger}(\omega) \right  \rangle   \, ,
\end{equation}
which is the eigenstate-free representation  of the XAS cross section. Taking matrix elements over the electronic degrees of freedom only (for simplicity the electronic states are also denoted by lowercase letters, which should not lead to any confusion since the context is different), and reinserting a resolution of the identity in the electronic final states, leads to the form that will be suitable for the semiclassical approximation
\begin{equation}
\sigma(\omega) \propto \frac{\Gamma_f}{\pi}  \sum_{f} \left \langle  D_{if}(\omega) D_{fi}^{\dagger}(\omega) \right  \rangle   \, .
\label{eq: XAS_quantum_final}
\end{equation}

\subsection{Time-domain formulation of RIXS}

When the system initially is in state $| i \rangle$, the double differential cross section where an incoming photon with frequency $\omega$ scatters to an outgoing photon with frequency $\omega'$ is \cite{Gelmukhanov_1999}
\begin{align}
&\frac{d^2 \sigma(\omega', \omega)}{d \omega d\Omega} = \frac{\omega'}{\omega} \sum_f |F_f(\omega)|^2\delta(\omega' -\omega + E_{fi}) 
\label{eq:KH} \\
&F_f(\omega) =  \alpha \sum_n  \frac{\langle f | D^{\dagger'} | n \rangle \langle n | D | i \rangle}{\omega - E_{ni} + i\Gamma} \, ,
\label{eq:KH_eigenstate_omp}
\end{align}
where the half-width-half-maximum broadening parameter $\Gamma$ is due to the electronic lifetime. Also the fine structure constant $\alpha$ is present. The transition operator $D$ depends on the incoming field and the primed $D'$ on the outgoing field. 
The KH formula in Eqs. (\ref{eq:KH},\ref{eq:KH_eigenstate_omp}) is essentially the one stated in Ref. \onlinecite{Gelmukhanov_1999}, but including only the resonant term. 

To connect to experiment, the double differential cross section should be convoluted with the incoming frequency distribution $\Phi_{inc}(\omega)$  and with a function describing the instrumental resolution $\Phi_{ins}(\omega')$ as 
\begin{equation}
\begin{split}
\sigma(\omega', \omega) =& \int^{\infty}_{-\infty} d \omega_0' \left [ \int^{\infty}_{-\infty} d \omega_0 \frac{d^2 \sigma(\omega_0', \omega_0)}{d \omega d\Omega} \Phi_{inc}(\omega-\omega_0) \right ] \\
& \times \Phi_{ins}(\omega'-\omega_0')  \, .
\label{eq:KH_convolution}
\end{split}
\end{equation}
To transform the RIXS cross section to the time domain the same steps are followed as for the XAS cross section. Using the representation of the delta function in Eq. (\ref{eq:delta_function}) 
the cross section becomes
\begin{equation}
\frac{d^2 \sigma(\omega', \omega)}{d \omega d\Omega} = \frac{\omega' \Gamma_f }{\omega \pi} \sum_f \left | \frac{F_f(\omega)}{\omega' -\omega + E_{fi} + i\Gamma_f} \right | ^2  \, ,
\label{eq:KH_square}
\end{equation}
where $\Gamma_f$ is small broadening parameter. 
Using Eq. (\ref{eq:resolvent}) the amplitude 
$F_f(\omega)$ 
can be transformed to the time domain
\begin{equation}
\begin{split}
F_f(\omega) &=  -i \alpha \int^{\infty}_{0} dt \sum_n  \langle f | D^{\dagger'} | n \rangle \langle n | D | i \rangle e^{i(\omega - E_{ni} + i\Gamma)t}\\
&= -i\alpha \int^{\infty}_{0} dt \sum_n  \langle f | D^{\dagger'} | n \rangle \langle n | e^{-iHt} D  e^{iHt}| i \rangle e^{i(\omega +i\Gamma)t} \\
&= -i\alpha \int^{\infty}_{0} dt \langle f |  D^{\dagger'}(0) D(-t) | i \rangle e^{i(\omega +i\Gamma)t} \, .
\end{split}
\end{equation}

The same procedure is performed for the denominator coming from the delta function to get
\begin{equation}
\begin{split}
\frac{F_f(\omega)}{\omega' -\omega + E_{fi} + i\Gamma_f}  = \langle f | F(\omega, \omega'-\omega)  | i \rangle
\end{split}
\end{equation}
with
\begin{equation}
\begin{split}
F(\omega,  \omega'-\omega) =& -\alpha \int^{\infty}_{0} dt \int^{\infty}_{0} dt'  D^{\dagger'}(t') D(t'-t) \\
&e^{i(\omega +i\Gamma)t} e^{i(\omega' -\omega+i\Gamma_f) t'} \, .
\end{split}
\end{equation}

Expanding the square in Eq. (\ref{eq:KH_square}) and using the resolution of the identity $\sum |f \rangle \langle f | =1$, the eigenstate-free representation of the Kramers-Heisenberg cross section is obtained as
\begin{equation}
\frac{d^2 \sigma(\omega', \omega)}{d \omega d\Omega} = \frac{\omega' \Gamma_f}{ \omega \pi}  \Big \langle   F^{\dagger}(\omega, \omega'-\omega)  F(\omega, \omega'-\omega) \Big \rangle \, .
\end{equation}

This formula is already generalized to a thermal average as was done for the XAS cross section. Proceeding to take electronic matrix elements yields
\begin{equation}
\frac{d^2 \sigma(\omega', \omega)}{d \omega d\Omega} =  \frac{\omega' \Gamma_f}{ \omega \pi}  \sum_f \left  \langle   F^{\dagger}_{if}(\omega, \omega'-\omega)  F_{fi}( \omega, \omega'-\omega) \right \rangle \, ,
\label{eq:KH_timedomain_final}
\end{equation}
with
\begin{equation}
\begin{split}
F_{fi}(\omega,  \omega'-\omega) =& -\alpha \sum_n \int^{\infty}_{0} dt \int^{\infty}_{0} dt'  D_{fn}^{\dagger '}(t') D_{ni}(t'-t) \\
&e^{i(\omega +i\Gamma)t} e^{i(\omega' -\omega+i\Gamma_f) t'} \, .
\label{eq:KH_time_states}
\end{split}
\end{equation}

If Eq. 
(\ref{eq:KH})
 is integrated over the incoming frequency $\omega$, which corresponds to a broadband incoming excitation the same development as before can be followed to obtain
\begin{align}
\langle \sigma(\omega') \rangle_{\omega} &\approx \frac{\Gamma}{\pi}  \sum_{f} \left \langle F^{\dagger \langle\omega\rangle}_{if}(\omega') F^{\langle\omega\rangle}_{if}(\omega') \right \rangle \, ,\\
F^{\langle\omega\rangle}_{if}(\omega')  &= -i \alpha  \sum_{n} \int_0^{\infty}dt' D_{fn}^{\dagger '}(t') D_{ni}(0)e^{i(\omega' +i\Gamma)t'} \, , 
\end{align}
which is proportional to the non-resonant XES cross section already used in \cite{Ljungberg_2010, Ljungberg_2011, Ljungberg_2017}. If instead the outgoing frequency $\omega'$ is integrated over the following expression is obtained 
\begin{align}
\langle \sigma(\omega) \rangle_{\omega'} &\approx \frac{\Gamma}{\pi}  \sum_{f} \left \langle F^{\dagger \langle\omega'\rangle}_{if}(\omega)  F^{\langle\omega'\rangle}_{if}(\omega) \right \rangle \\
F^{\langle\omega'\rangle}_{if}(\omega)   &= -i \alpha  \sum_{n} \int_0^{\infty}dt' D_{fn}^{\dagger '}(0) D_{ni}(-t)e^{i(\omega +i\Gamma)t} 
\end{align}
which except for the extra dipole operator is close to the XAS cross section in Eq. (\ref{eq: XAS_quantum_final}). The RIXS cross section averaged over either the incoming or the outgoing frequencies are important special cases that should be well reproduced in the semiclassical method.

\subsection{Polarization dependence}

For randomly oriented molecules in gases and and liquids the cross section should be orientationally averaged with respect to the polarization direction of incoming and outgoing electric fields. The polarization dependence of the RIXS cross section is discussed detail in Refs. \onlinecite{Luo_1994, Luo_1996, Gelmukhanov_1994} where also explicit formula are provided in terms of the Cartesian directions of the transition dipoles. The formulae for linearly polarized light is easily adapted to the eigenstate-free representation of RIXS as
\begin{equation}
\begin{split}
\langle \sigma(\omega', \omega) \rangle_{\text{orient}} =&  \frac{\omega' \Gamma_f}{ \omega \pi}  \sum_f  (-\lambda^F_f + 4\lambda^G_f  -\lambda^H_f) \\
&+(3 \lambda^F_f - 2\lambda^G_f  -3\lambda^H_f)\cos^2\theta
\end{split}
\label{eq:pol_dep_theta}
\end{equation}
with $\theta$ the angle between the polarization direction of the incoming and outgoing polarization directions and
\begin{align}
\lambda^F_f &=&  \sum_{\beta, \gamma} \left \langle F^{\beta \beta \dagger}_{if}(\omega, \omega'-\omega)  F^{\gamma \gamma}_{fi}(\omega, \omega'-\omega) \right \rangle \, ,\\
\lambda^G_f &=&  \sum_{\beta, \gamma} \left \langle F^{\beta \gamma \dagger}_{if}(\omega, \omega'-\omega)  F^{\beta \gamma}_{fi}(\omega, \omega'-\omega) \right \rangle \ , \\
\lambda^H_f &=&  \sum_{\beta, \gamma} \left  \langle F^{\beta \gamma \dagger}_{if}(\omega, \omega'-\omega)  F^{\gamma \beta}_{fi}(\omega, \omega'-\omega) \right \rangle \, .
\end{align}

The greek indices denote Cartesian components of the electric field in the dipole operators appearing in the scattering amplitude
\begin{equation}
\begin{split}
F^{\beta\gamma}_{fi}(\omega,  \omega'-\omega) =& -\alpha \sum_n \int^{\infty}_{0} dt \int^{\infty}_{0} dt'  D_{fn}^{\gamma \dagger '}(t') D^{\beta}_{ni}(t'-t) \\
&e^{i(\omega +i\Gamma)t} e^{i(\omega' -\omega+i\Gamma_f) t'} \, .
\label{eq:KH_time_states_pol}
\end{split}
\end{equation}

In the same way also expressions for circularly and elliptically polarized light can be adapted to the present scheme. When developing the semiclassical approximations the polarization dependence will not be explicitly written out in order not to complicate the description, the above formulae will be unchanged for these cases. For completeness the orientally averaged XAS cross section is also written out
\begin{equation}
\langle \sigma(\omega) \rangle_{\text{orient}} \propto \frac{\Gamma_f}{3\pi}  \sum_{f \beta} \left \langle  D^{\beta}_{if}(\omega) D_{fi}^{\beta \dagger}(\omega) \right  \rangle   \, .
\label{eq: XAS_quantum_final}
\end{equation}

\subsection{Semiclassical time evolution of operators}

In this section an approximate description of the time evolution of operators will be described. The formalism is close to semiclassical schemes for other spectroscopies due to Oxtoby \cite{Oxtoby_1978} and others \cite{Mukamel_1982, Mukamel_book, Auer_2007} and is derived with aim to include the possibility of non-adiabatic effects (although in the present publication this feature is not explored further). 

The time evolution of an operator $O$ in the Heisenberg picture is 
\begin{equation}
O(t) = U^{\dagger}(t, t_0) O(t_0)  U(t, t_0)  \, ,
\end{equation}
where the time evolution operator $U(t,t_0)$ depends on the full Hamiltonian $H$ of the system as
\begin{equation}
U(t, t_0) = e^{-iH(t-t_0)} \, .
\end{equation}
In the Born-Oppenheimer approximation (that here is assumed to be valid for the ground state, but not necessarily for excited states), a wave function that describes both electronic and nuclear degrees of freedom can be written as
\begin{equation}
\Psi(\mathbf{r},\mathbf{R}) = \chi(\mathbf{R})   \psi(\mathbf{r};\mathbf{R}) \, ,
\label{eq:wfn_BO}
\end{equation}
where $ \psi(\mathbf{r};\mathbf{R})$ is a solution of the electronic Hamiltonian with fixed nuclear coordinates $\mathbf{R}$, and $ \chi(\mathbf{R})$ is the vibrational wave function. It is in principle possible to relax the Born-Oppenheimer approximation even for the ground state by replacing Eq. (\ref{eq:wfn_BO}) by a sum over states, but this will not be considered in the following.
Now a classical time-evolution operator $\widetilde U^{l}(t,t_0)$Ê is introduced that has the effect
\begin{equation}
\widetilde U^{l}(t,t_0)  \chi(\mathbf{R}(t_0))   \psi(\mathbf{r};\mathbf{R}(t_0)) = \chi(\mathbf{R}(t_0))   \psi(\mathbf{r};\mathbf{R}(t)) \, ,
\end{equation}
that is to map the positions of the nuclei $\mathbf{R}(t_0)$ at time $t_0$ to some other positions $\mathbf{R}(t)$ in $\psi(\mathbf{r};\mathbf{R})$, and let the electronic degrees of freedom follow this change adiabatically. The coefficients $\chi(\mathbf{R})$ are not touched by the operation. 
The dynamics of the nuclear coordinates is assumed to take place classically on a  potential energy surface denoted by the superscript $l$. The classically evolved operator, denoted with a $\sim$, is then
\begin{equation}
\widetilde O(t) = \widetilde U^{l \dagger}(t, t_0) O(t_0)  \widetilde U^l(t, t_0) \, .
\end{equation}
If the quantum operators are naively replaced with the classical ones, an expectation value would look like
\begin{equation}
\langle \Psi_i | \widetilde O(t) | \Psi_i \rangle  = \int dR |\chi_i(\mathbf{R})|^2  
\widetilde O_{ii}(t) \, ,
\label{eq:class_expectation value}
\end{equation}
with
\begin{equation}
\widetilde O_{ii}(t) = \int dr \psi^*_i(\mathbf{r},\mathbf{R}) \widetilde O(t) \psi_i(\mathbf{r},\mathbf{R}) \, .
\end{equation}
that is an average over the ground state distribution $|\chi_i(\mathbf{R})|^2$ of the classically evolved matrix elements. This expression 
arises as
a consequence of the fact that $\chi_i(\mathbf{R})$ is not touched by the classical time development. A generalization of Eq. (\ref{eq:class_expectation value}) to include an ensemble average over the vibrational states is straightforward and will only affect the distribution $|\chi_i(\mathbf{R})|^2$. A purely classical approach misses out on all quantum effects, in particular quantum transitions. However, it is possible to include some of these effects in a less naive semiclassical scheme that still retains the simple average in  Eq. (\ref{eq:class_expectation value}).

The classical as well as the quantum time-evolution operators are unitary, that is $U^{\dagger}(t,t_0) U(t,t_0) = U(t,t_0) U^{\dagger}(t,t_0) =1$. This leads to the exact representation 
\begin{equation}
O(t) = \bar U^{l \dagger}(t, t_0) \widetilde O(t) Ê\bar U^{l}(t,t_0) \, ,
\label{eq:time_evolution_mixed}
\end{equation}
where the barred "mixed" time evolution operator is
\begin{equation}
\bar U^{l}(t, t_0)  = \widetilde U^{l\dagger}(t,t_0) U(t,t_0) \, .
\label{eq:quantum_time_evolution}
\end{equation}
This representation decomposes the dynamics of an operator into a classical part, multiplied with quantum correction factors (mixed time-evolution operators) that can later be approximated. The operator $\bar U^{l}(t, t_0)$ is obtained using its equation of motion
\begin{equation}
\begin{split}
\frac{d}{dt} \bar U^{l}(t, t_0) &=  \left (\frac{d}{dt} \widetilde U^{l\dagger}(t,t_0) \right ) U(t,t_0) \qquad \\
& \quad ~ +  \widetilde U^{l\dagger}(t,t_0) \left ( \frac{d}{dt}U(t,t_0)\right)  \\
& =  \Bigg [ \left (\frac{d}{dt} \widetilde U^{l\dagger}(t,t_0) \right ) \widetilde U^l(t,t_0) \\
& \quad ~ - i\widetilde U^{l\dagger}(t,t_0) H \widetilde U^l(t,t_0) \Bigg ]\bar U^{l}(t, t_0) \, .
\end{split}
\label{eq:combined_time_evolution}
\end{equation}
The last identity was obtained by the time derivative of Eq. (\ref{eq:quantum_time_evolution}) and the insertion of $\widetilde U(t,t_0) \widetilde U^{\dagger}(t,t_0) =1$. Now matrix elements are taken using the states at $t_0$, that are time-independent solutions of the electronic Hamiltonian $H^{\text{el}} | a (\mathbf{r}; \mathbf{R}(t_0)) \rangle = E_a(\mathbf{R}(t_0)) | a (\mathbf{r}; \mathbf{R}(t_0)) \rangle $. For general states the letters a, b, c etc. are used and for simplicity the explicit dependence on $\mathbf{r}$ and $\mathbf{R}$ will be omitted  in the following. Eq. (\ref{eq:combined_time_evolution}) can then be written as 
\begin{equation}
\begin{split}
\frac{d}{dt} \bar U^{l}_{ab}(t, t_0) &=  -\sum_{c} \Big [ \langle a(t) | \frac{d}{dt} c(t) \rangle
  +i \langle a(t) | H | c(t) \rangle \Big ]\bar U^{l}_{cb}(t, t_0) \, .
\end{split}
\label{eq:bar_dUdt_ab}
\end{equation}
Here the relation $\widetilde U^l(t,t_0)  | a(t_0) \rangle =   | a(t) \rangle$ was used, as well as a partial integration to move the time derivative to the right. The initial condition for the differential equation is $\bar U^{l}_{cb}(t_0, t_0) = \delta_{cd}$. The first term in Eq. (\ref{eq:bar_dUdt_ab}) are the non-adiabatical coupling matrix elements. They appear because the instantaneous electronic states, when the nuclei are evolved classically, are not in general orthogonal. 
The full Hamiltonian $H=H^{\text{el}} + T^{\text{nuc}}$  contains both the electronic part $H^{\text{el}}$ and the nuclear kinetic energy operator $T^{\text{nuc}}$. The latter is difficult to treat as it would affect the initial vibrational wave function $\chi(\mathbf{R})$, so including it would not lead to a simple average as in Eq. (\ref{eq:class_expectation value}). For this reason it will be ignored and only the electronic Hamiltonian will be used. Then Eq. (\ref{eq:bar_dUdt_ab}) reduces to
\begin{equation}
\begin{split}
\frac{d}{dt} \bar U^{l}_{ab}(t, t_0) &=  -\sum_{c} \Big [ \langle a(t) |\frac{d}{dt}  c(t) \rangle
  +i E_a(t) \delta_{ac} \Big ]\bar U^{l}_{cb}(t, t_0) \, .
\end{split}
\label{eq:bar_dUdt_ab_general}
\end{equation}
When $\bar U^{l}_{ab}(t, t_0) $ has been determined from Eq. (\ref{eq:bar_dUdt_ab_general}) 
 the matrix elements of the operator $O(t)$ can be written as
\begin{equation}
O_{ab}(t) = \sum_{cd}  \bar U^{l \dagger}_{ac}(t, t_0)  \widetilde O_{cd}(t)  \bar U^{l}_{db}(t, t_0) \, .
\label{eq:eq_of_motion_adiabatic_solution}
\end{equation}
A further simplification of the problem is to assume that the non-adiabatic coupling matrix elements are zero, that is to do the
adiabatic approximation. This approximation turns the matrix equation of Eq. (\ref{eq:bar_dUdt_ab_general}) into a scalar equation
\begin{equation}
\begin{split}
\frac{d}{dt} \bar U^{l}_{aa}(t, t_0) &=  -i E_a(t) \bar U^{l}_{aa}(t, t_0) \, ,
\end{split}
\label{eq:eq_of_motion_adiabatic}
\end{equation}
with all non-diagonal terms zero because of the initial condition and the absence of couplings. Eq. (\ref{eq:eq_of_motion_adiabatic}) can be solved to yield
\begin{equation}
\bar U^{l}_{aa}(t, t_0) =  e^{-i \int_{t_0}^t d\tau E_a(\tau) } \, ,
\label{eq:eq_of_motion_adiabatic_solution}
\end{equation}
which gives the adiabatic  time evolution of an operator as 
\begin{equation}
O_{ab}(t) =  \widetilde O_{ab}(t)  e^{-i \int_{t_0}^t d\tau E_{ba}(\tau)} \, .
\label{eq:O_adiabatic}
\end{equation}
It should be noted that the actual potential energy surface where the dynamics is performed does not appear explicitly in Eq. (\ref{eq:O_adiabatic}), thus there is  freedom to choose where the time development takes place. In the adiabatic quantum case the matrix element is
\begin{equation}
O_{ab}(t) =  e^{iH_{a} (t-t_0)} O_{ab}(t_0) e^{-iH_{b} (t-t_0)} 
\end{equation}
which, operated onto a wave function on the right has the following interpretation: propagate the wave function on the PES $H_{b}$, then apply the operator $O(t_0)$ and then propagate back in time from time $t$ to $t_0$ on PES $H_a$. It is clear that such a matrix element inherently depends on two distinct PESes, and to approximate it to a single one makes a choice necessary. 

The initial conditions for the positions are given by the ground state vibrational distribution but the momenta are still free parameters. In Ref. \onlinecite{Ljungberg_2010} it was shown that sampling the momenta from the ground state momentum distribution gives the right spread of the time-dependent position distribution compared to full wave packet dynamics, and this approach will be used in the present publication.

\subsection{Semiclassical description of XAS and RIXS}

The semiclassical approximation to the XAS and RIXS cross sections can be developed using the results in the previous section. First XAS will be treated. Inserting the semiclassical approximation Eq. (\ref{eq:time_evolution_mixed}) of the time-dependent quantity $D_{if}(t)$ in 
Eq. (\ref{eq:D_omega_FT})
leads to the following expression
 \begin{equation}
D_{if}(\omega) =  -i\sum_{n,n'} \int^{\infty}_{0} dt  \bar U_{in}^{l \dagger}(t) \widetilde D_{nn'}(t) Ê\bar U_{n',f}^{l}(t) e^{i(\omega +i\Gamma) t} \, ,
\end{equation}
or, using the adiabatic approximation
 \begin{equation}
D_{if}(\omega) =   -i\int^{\infty}_{0} dt \widetilde D_{if}(t) e^{-i \int_{0}^t d\tau E_{fi}(\tau) }  e^{i(\omega +i\Gamma) t} \, ,
\end{equation}
where $\widetilde D_{if}(t)$ is the classically developed matrix element of the dipole operator and $E_{fi}(\tau)$ are the corresponding state energies.
Now the classical Franck-Condon approximation is introduced. It amounts to assuming that for a given value of the nuclear coordinate $\mathbf{R}$ the transition frequency is constant, that is, one goes directly from one point of the ground state PES to the excited state PES. Then $E_{fi}(t) = E_{fi}(\mathbf{R})$  and $\widetilde D_{if}(t) = \widetilde D_{if}(\mathbf{R})$, which leads to

 \begin{equation}
 \begin{split} 
D_{if}(\omega, \mathbf{R}) &=   -i \int^{\infty}_{0} dt \widetilde D_{if}(\mathbf{R}) e^{-i E_{fi}(\mathbf{R}) t }  e^{i(\omega +i\Gamma) t} \\
&= \frac{\widetilde D_{if}(\mathbf{R})}{\omega - E_{fi}(\mathbf{R}) + i \Gamma} \, .
\, 
\end{split}
\end{equation}

The vibrational effects in this approximation come from the sampling of the $\mathbf{R}$ coordinates according to the initial state vibrational distribution. Quantum mechanically, this approximation gives the right limit for a dissociative final state that can be described as being linear in the region of the initial state distribution \cite{Gelmukhanov_1999}. The limiting case with dissociative behavior is important to describe well since in application this will be a common situation (and will be where most vibrational effects occur). An additional improvement of the classical FC approximation is to include a zero-point energy correction, that is $E_i(\mathbf{R}) = E_{ZPE}$, which is justified since for an initial vibrational state in the ground state the same energy is excited from, independent of where on the PES $\mathbf{R}$ is located. In Ref. \onlinecite{Leetmaa_2010} the classical FC with and without ZPE correction was used to compute the FC profile of XAS of a water molecule, and the ZPE seemed to improve the description there. Note that the classical FC approximation does not require explicit nuclear dynamics, and so will be quite a lot cheaper than the original semiclassical approximation. Since it also has the correct quantum limit in the dissociative case, it could even give better results in certain circumstances. 

Since XAS involves a single time argument it is much simpler than RIXS, which involves two. 
According to Eq. (\ref{eq:KH_time_states}) the time argument $t'$ describes the emission, while $t'-t$ should describe the absorption as well as the connection between absorption and emission. Since the non-resonant XES has been shown to work very well when the dynamics is performed on the dissociative intermediate state, it seems reasonable to keep the time evolution for $t'$ to be on an intermediate state, however, in RIXS there is an additional issue in that there are possibly several intermediate electronic states and it is not really clear from the outset which one to use for the dynamics. For the other time argument $t'-t$ it is not obvious what to do; on one hand it should describe absorption, so it should involve geometries not too far from the equilibrium geometry, on the other hand it must also connect the absorption and emission processes which means that it should also describe the dissociation. 
Using the adiabatic approximation the semiclassical expression for the RIXS cross section reads
\begin{equation}
\begin{split}
F_{fi}(\omega,  \omega'-\omega) =& -\alpha \sum_n \int^{\infty}_{0} dt \int^{\infty}_{0} dt'  \widetilde D_{fn}^{\dagger '}(t') \widetilde D_{ni}(t'-t) \\
& e^{-i\int_0^{t'} d\tau E_{nf}(\tau) } e^{i\int_0^{t'-t}  d\tau E_{ni}(\tau)} \\
&e^{i(\omega +i\Gamma)t} e^{i(\omega' -\omega+i\Gamma_f) t'} \, .
\label{eq:SCKH_RIXS_adiabatic}
\end{split}
\end{equation}

One way to simplify this expression is to use different approximations for the absorption end emission processes. Using the classical FC approximation for $\widetilde D_{ni}$ (the absorption) results in  $e^{i\int_0^{t'-t} d\tau E_{ni}(\tau) }
\approx e^{i E_{ni}(\mathbf{R})t'} e^{-i E_{ni}(\mathbf{R})t}$ and $\widetilde D_{ni}(t'-t) \approx \widetilde D_{ni}(\mathbf{R})$, and this leads to
\begin{equation}
\begin{split}
F_{fi}(\omega,  \omega'-\omega, \mathbf{R}) =& -\alpha \sum_n \int^{\infty}_{0} dt  \widetilde D_{ni}(\mathbf{R}) e^{i (\omega - E_{ni}(\mathbf{R}) +i\Gamma)t} \\
& \cdot \Big ( \int^{\infty}_{0} dt'
\widetilde D_{fn}^{\dagger '}(t')
e^{-i\int_0^{t'} d\tau E_{nf}(\tau) } \\
&e^{-i(\omega -E_{ni}(\mathbf{R}))t'} 
e^{-\Gamma_f t'} e^{i\omega'  t'} \Big ) \, .
\end{split}
\end{equation}

Integrating over $t$ this becomes
\begin{equation}
\begin{split}
F_{fi}(\omega,  \omega'-\omega, \mathbf{R}) =& -\alpha \sum_n  \frac{\widetilde D_{ni}(\mathbf{R})}{\omega - E_{ni}(\mathbf{R}) + i \Gamma}\\
& \cdot \Big ( \int^{\infty}_{0} dt'
\widetilde D_{fn}^{\dagger '}(t')
e^{-i\int_0^{t'} d\tau E_{nf}(\tau) } \\
&e^{-i(\omega -E_{ni}(\mathbf{R}))t'} 
e^{-\Gamma_f t'} e^{i\omega'  t'} \Big ) \, .
\label{eq:SCKH_RIXS}
\end{split}
\end{equation}

It is instructive to compare Eq. (\ref{eq:SCKH_RIXS}) to the expression for the non-resonant (or ionized case) presented in Ref \onlinecite{Ljungberg_2010}. In this case there is always resonance in the absorption, as a continuum electron is ejected that can take up any energy. Then Eq. (\ref{eq:SCKH_RIXS}) reduces to $ \widetilde D_{ni}(\mathbf{R}) \int^{\infty}_{0} dt' \widetilde D_{fn}^{\dagger '}(t') e^{-i\int_0^{t'} d\tau E_{nf}(\tau) } e^{-\Gamma_f t'} e^{i\omega'  t'}$ which means that $\Gamma_f$ turns up where the lifetime broadening $\Gamma$ should be. This fact is due to the different approximations for the absorption and the emission in the current scheme. It turns out that setting $\Gamma_f = \Gamma$ "corrects" this shortcoming, and this will be done in the following. In the quantum case the correction corresponds to an additional Lorentzian instrumental broadening with HWHM $\Gamma$ that should be relatively unproblematic since the instrumental broadening usually is larger than the lifetime broadening.  

\section{Results and discussion}

\subsection{Harmonic test system for XAS} 

First the semiclassical approximation to XAS will be investigated for different one-dimensional potential energy surfaces. In order to have a set of models that go from a bound to a dissociative excited state, harmonic potentials with shifting frequency, center and depth are used. The value of the excited-state potentials and their derivative is constrained to be equal at  $R_0$, the center of the ground state potential. The ground state potential $V_0(R_0)$ is modeled as a harmonic potential with a frequency corresponding roughly to a hydrogen stretch mode ($\sim$ 3500 cm$^{-1}$)
\begin{equation}
V_0(R) = \frac{1}{2} \mu \omega_0^2(R-R_0)^2 + V^0_0 \, ,
\end{equation}
and the excited state potentials are taken as
\begin{equation}
V_n(R) = \frac{1}{2} \mu \omega_n^2(R-R_n)^2 + V^0_n \, ,
\end{equation}
where for given $R_0$, $R_1$, $R_n$, $V^0_1$ and $\omega_1$ the needed parameters are
\begin{eqnarray}
\omega_n^2 &=&  \omega_1^2 \left( \frac{R_0-R_1}{R_0-R_n} \right ) \, ,\\
V^0_n &=& \frac{1}{2} \mu  \omega_1^2 (R_0-R_1)(R_n-R_1) + V^0_1 \, .
\end{eqnarray}
The chosen parameters are $R_0 =1$ \AA,  $R_1 = R_0 +0.1$ \AA,  and in general $R_n = R_0 + n \cdot 0.1$ \AA \, and  $\omega_1 = \omega_0$. This choice 
leads to a first excited state that has the same frequency as the ground state potential, only with the position slightly shifted. Thus the exited state frequencies will go from about 3500 to 1000 wave numbers, the latter for the most displaced and dissociative state. 
PESes with high value of the parameter $n$ will more and more approach a dissociative potential. $V^0_1$ is arbitrarily chosen to be 100 eV. %
The potential energy surfaces are shown in the Supplementary Material \cite{Sup_Mat}. The vibrational states are solved using a Fourier grid method with a grid spacing of 0.01 \AA. For the semiclassical method trajectories are run for 102.4 fs in steps of 0.05 fs using the velocity Verlet algorithm. A lifetime broadening of 0.045 eV is used with an instrumental broadening of 0.1 eV. The initial conditions are sampled uniformly with 20 position and 20 momentum points.  

\begin{figure}[ptb]
\includegraphics[width=1.0\columnwidth, angle=0]{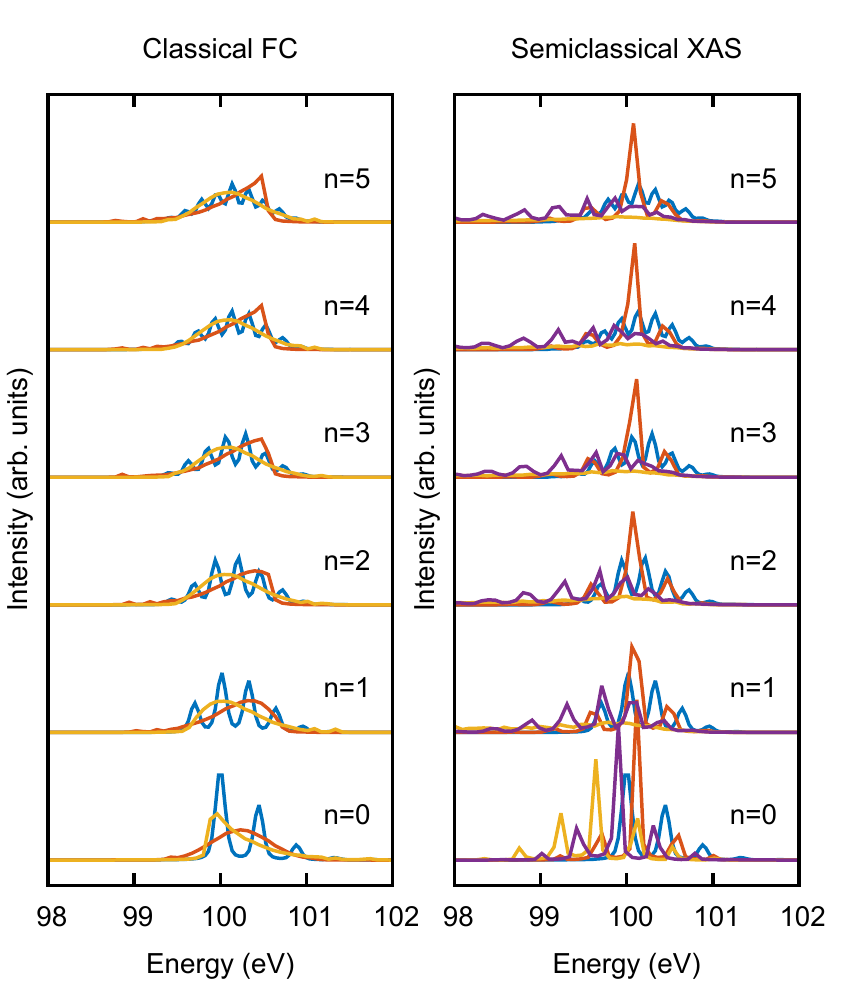}
 \caption{XAS spectra for the high-frequency ground state. Left: the classical FC method. Blue:  quantum XAS, red: classical FC without ZPE, yellow: classic FC with ZPE. Right: semiclassical XAS. Blue:  quantum XAS, red: dynamics on ground state, yellow: dynamics on excited state, purple: dynamics on the average between ground and excited states. The spectra are normalized.}
\label{fig:xas_FC_high}
\end{figure}

In Fig. \ref{fig:xas_FC_high} different approximations of the XAS spectra are shown for the six PESes, were the spectrum at the bottom  is the one with highest frequency PES and the top one is the one with the lowest. For the quantum case a clear FC progression is visible that is asymmetric for the more bound PES and more and more symmetric with closely spaced vibrational peaks for the more dissociative PESes. Indeed, the limit for a fully dissociative state should be a Gaussian distribution. The classical FC approximation does not resolve the vibrational states at all. Without  the ZPE correction the line shape goes from approximately Gaussian for the most bound state to an asymmetric spectrum towards high energies for the more distorted states, clearly in contrast to the quantum spectrum. When the ZPE correction is introduced, however, the line shape follows the envelope of the quantum spectrum well. 

In the semiclassical  XAS the dynamics can be run either on the ground state PES or on the excited state PES, or possibly on an average of the two. In Fig. \ref{fig:xas_FC_high} these approximations are shown together with the quantum case. The semiclassical scheme does capture some vibrational structure of the spectrum, but it is not really close to the quantum one. For dynamics on the ground state the spectrum changes little depending on the excited state PES, and just broadens a bit with more distorted final states. For dynamics on the final state the spectra become very smeared out and go too much to low frequencies, and for the averaged PES the same happens, but to a lesser degree.

\begin{figure}[ptb]
\includegraphics[width=1.0\columnwidth, angle=0]{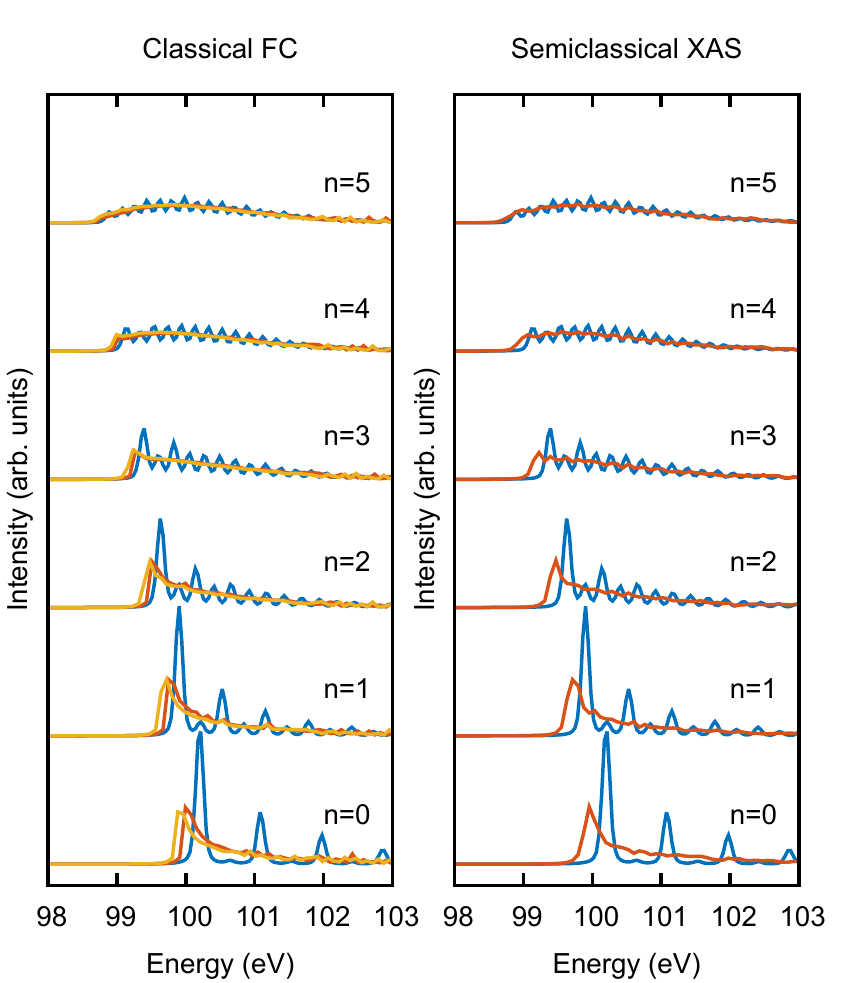}
 \caption{XAS spectra for the low-frequency ground state. Left: the classical FC method. Blue:  quantum XAS, red: classical FC without ZPE, yellow: classic FC with ZPE. Right: semiclassical XAS. Blue:  quantum XAS, red: dynamics on ground state}
 \label{fig:xas_FC_low}
\end{figure}

To test the model for a case where the ground state PES has a lower vibrational frequency this PES is modified to have ten times lower frequency (350 cm$^{-1}$), with all the other PESes kept the same. In Fig. \ref{fig:xas_FC_low} the results are shown for this case. Here the FC overlap gives more vibrational structure since the ground state PES is more different from the excited state PESes than in the previous case. The classical FC, with or without the ZPE captures the envelope well, which also is the case for the semiclassical XAS where the dynamics is run on the ground state (the dynamics on the final state and the averaged one are not shown since they give bad results also in this case).   

In summary, the classical FC with ZPE correction seems to be a well-balanced approximation that works reasonably well for all model cases, indeed it is the only one of the considered methods that gives a good agreement with the quantum spectrum for the cases of a high-frequency ground state. When the ground state has a low vibrational frequency all the methods give satisfactory agreement. This suggests that the absorption part in the RIXS process could be treated with the classical FC approximation, and that the ZPE should be considered for the high-frequency modes.  

\subsection{RIXS for the L-shell excitation of HCl}

As a clear-cut test system for dissociative behavior on a core-excited state the L-edge core excited HCl molecule is chosen. This system was studied in work by Salek et al. \cite{Salek_1999, Salek_thesis} from which PESes for the initial, intermediate and final states were obtained. The PESes
show a clear dissociation for the intermediate (core excited) state, and for distances of more than 4 \AA \, the potentials are flat as the atoms are completely separated
(see the Supplementary Material  \cite{Sup_Mat}).
In Fig.  \ref{fig:KH_res_HCl} RIXS maps using KH (upper) and SCKH (lower) are shown.
In the case of KH the spectrum is computed according to Eqs. (\ref{eq:KH}, \ref{eq:KH_eigenstate_omp}) with the vibrational states  solved with a Fourier grid method on a real space grid ranging from 0.8 to 30.0 \AA , with a grid spacing of 0.0146 \AA. The transition dipoles are set to unity. A HWHM lifetime broadening of 0.045 eV is used with an incoming broadening of 0.3 eV and a instrumental broadening of 0.1 eV (both Gaussian). Additionally a Lorentzian instrumental broadening of 0.045 eV is applied to compensate for the corrected lifetime in the SCKH approach.
The sharp peak at around 181 eV is a clear indication of the dissociation, and corresponds to the energy separation of the intermediate and final state PESes when dissociated. This feature does not move with the incoming frequency. The other clear feature is the peak at around 185 eV that moves linearly with the incoming energy, and corresponds to molecular features, i. e. emission close to the ground state equilibrium distance. 
For the SCKH method, the spectrum is computed using Eqs. (\ref{eq:KH_timedomain_final}, \ref{eq:SCKH_RIXS}) with the dynamics run in steps of 0.1 fs for 102.4 fs. To avoid the classical trajectories falling of the edge of the PES, "absorbing boundary conditions" are used, meaning that when a trajectory reaches 20 \AA, it is considered to stay at that point at all later times. A number of 400 trajectories are used with starting conditions evenly sampled from the  position and momentum ground state quantum distributions (40x10 points respectively) that are obtained from the first vibrational eigenstate.
The main reason for the fine sampling of the position distribution is to build up the XAS profile in the classical FC scheme. A lower sampling can be compensated by introducing a Gaussian broadening function with a larger width. 
The same lifetime and broadening parameters were used as in the KH case, except for the additional Lorentizian instrumental broadening that is included as an artificial lifetime in the SCKH method. 

The general shape of the spectrum reproduces the KH one well, 
although
the molecular feature is too weak 
It is slightly dispersing linearly by the proper amount, but sharpness is missing. Comparing the colormaps the absorption and emission seem to occur at more or less the same frequencies for both methods. For actual modeling of i. e. liquids, such dissociation is quite extreme (instead, a more particle-in-a-box type situation is more commonly found) and it is thus comforting that the SCKH method is able to capture the right features qualitatively, and almost quantitatively.  

\begin{figure}[ptb]
\includegraphics[width=1.0\columnwidth, angle=0]{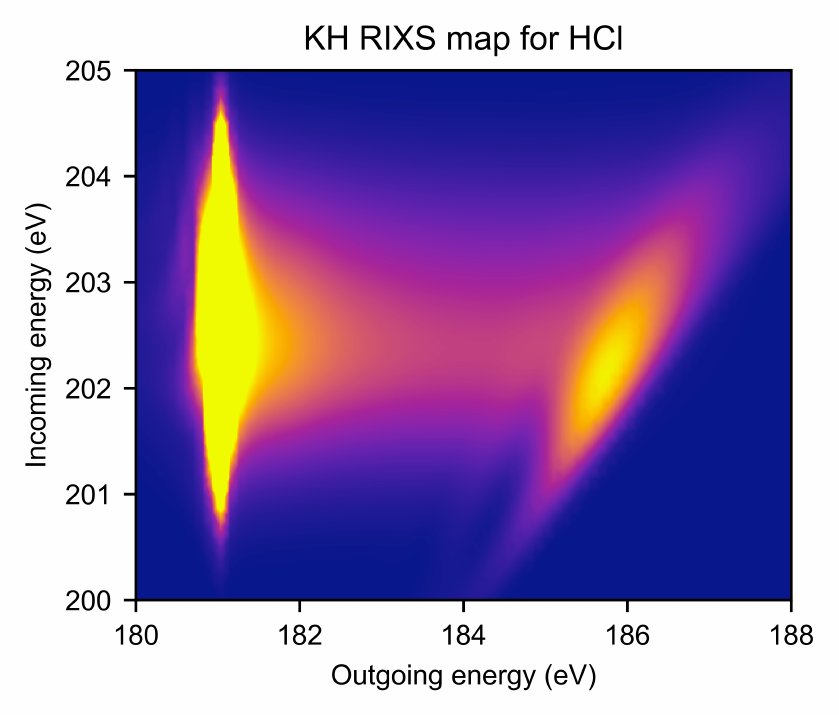}
\includegraphics[width=1.0\columnwidth, angle=0]{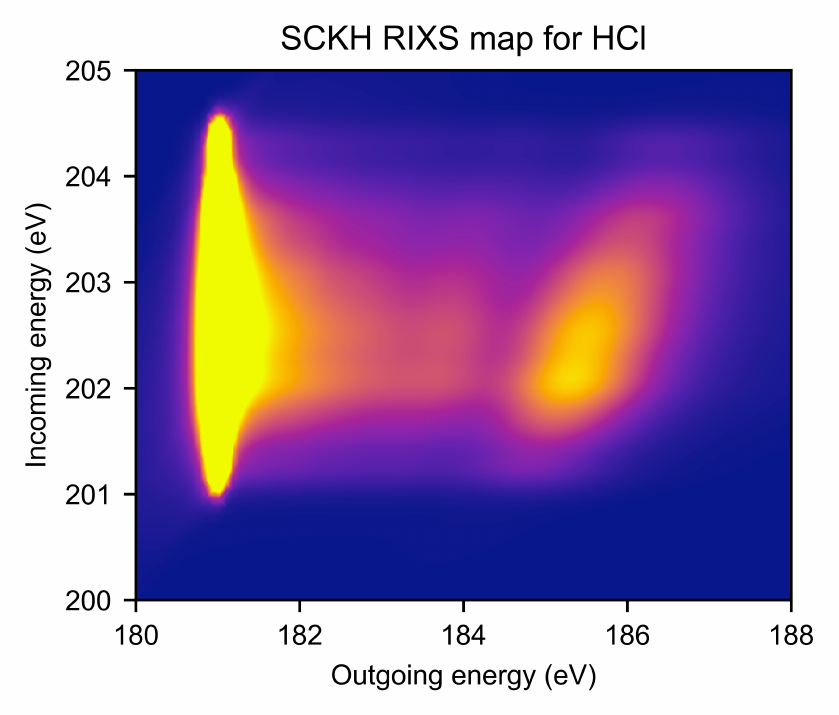}
 \caption{RIXS maps for HCl. Upper: KH, lower: SCKH.}
\label{fig:KH_res_HCl}
\end{figure}

\begin{figure}[ptb]
\includegraphics[width=1.0\columnwidth, angle=0]{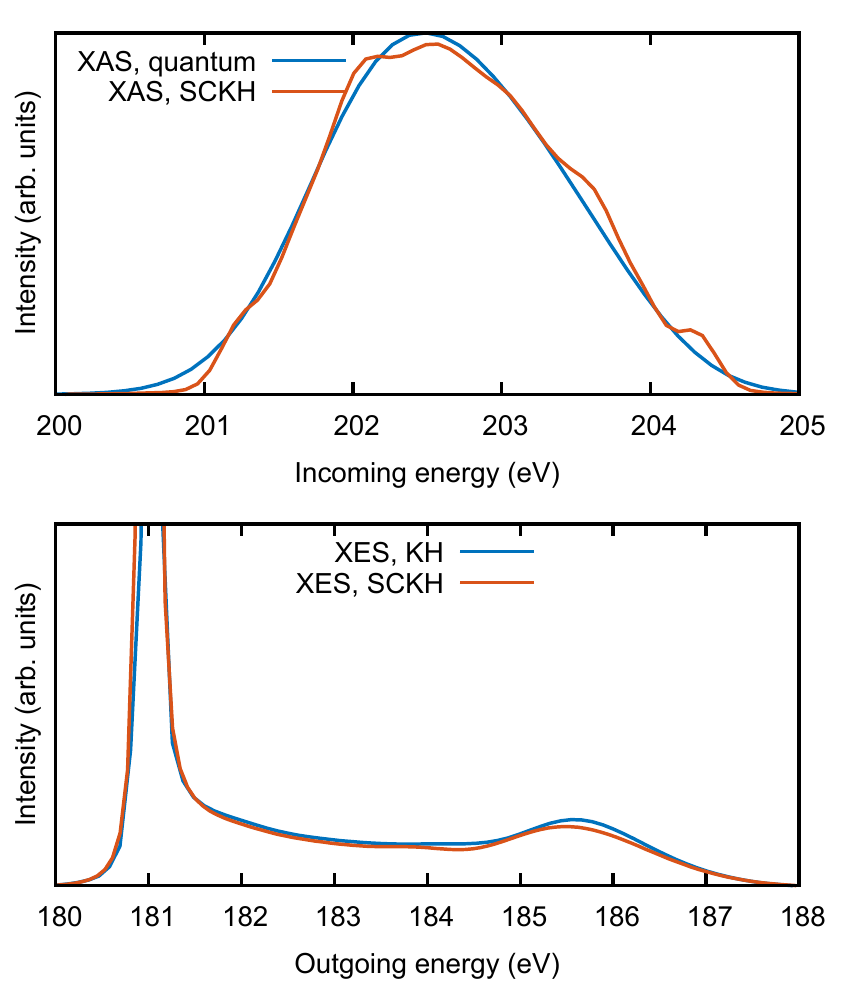}
 \caption{Averaged spectra for HCl. Upper frame: averaged spectrum over outgoing frequency. Blue: KH, red: SCKH. Lower frame: spectrum averaged over incoming frequency. Blue: KH, red: SCKH. The spectra are normalized.}
\label{fig:SCKH_res_HCl}
\end{figure}

In Fig. \ref{fig:SCKH_res_HCl} the averaged cross sections $\langle \sigma(\omega) \rangle_{\omega'}$, proportional to XAS, and $\langle \sigma(\omega') \rangle_{\omega}$ corresponding to broadband XES are shown for the KH and SCKH methods. The agreement is excellent, and in the case of $\langle \sigma(\omega) \rangle_{\omega'}$ the absorption profile due to the classical FC approximation is mostly dependent on the sampling on the x-coordinate, and can be even more converged if needs be.  

\begin{figure}[ptb]
\includegraphics[width=1.0\columnwidth, angle=0]{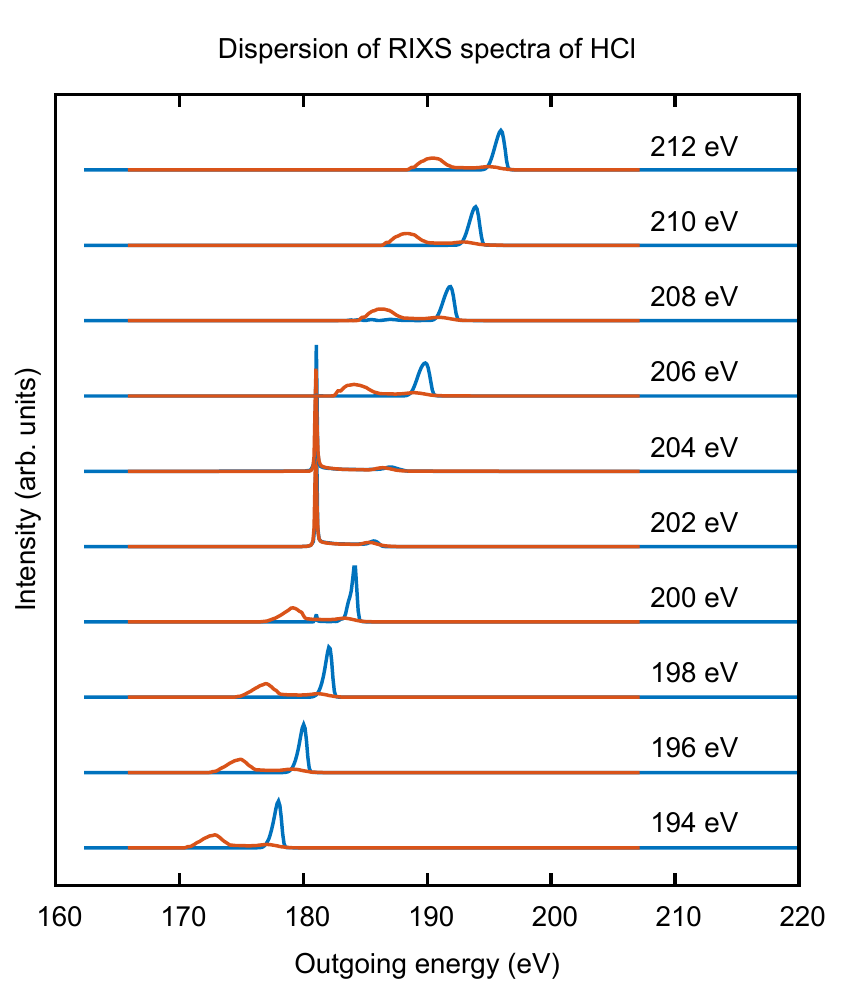}
 \caption{Normalized emission spectra for HCl. Blue: KH, red: SCKH. The absorption energy goes from 194 eV to 212 eV, bottom to top. }
\label{fig:dispersion_HCl}
\end{figure}

In RIXS, the spectrum can be approximately decomposed into a static part that is independent of the incoming energy and a part whose emission frequency disperses linearly with the incoming energy. The static part is apparently well-described by the SCKH formalism but it would also be of interest to investigate the dispersing part. In Fig. \ref{fig:dispersion_HCl} a comparison is made between the normalized emission spectra at a given incoming energy for the KH and SCKH cases. The KH spectra show a rather narrow dispersing peak, and when it hits the region of the molecular (non-dissociated) emission feature resonance occurs and the spectrum then resembles the averaged one. After the resonance region the peak disperses to higher energies. For the SCKH case, the dispersing feature instead looks like a broadened copy of the emission spectrum at resonance, with its dissociative feature coming out strongly. In the resonance region the static spectrum dominates and the agreement to the quantum results is again good. So there is a significant error of the method with respect to the dispersive part that one should be cautious about. 

\subsection{RIXS of a model methanol dimer}

Since the main motivation for this work is to model RIXS of liquids, the next model system considered is a simplification of a common situation that will occur in liquid methanol. The same model of a one-dimensional methanol dimer at the ground state equilibrium distance as in Ref. \onlinecite{Ljungberg_2017} is used, but with the difference that several intermediate states are computed with final states corresponding to each one. This is done approximatively by explicitly computing the first core- excited state (with a core hole and an excited election), and defining the higher ones by a one-electron excitation from this reference state using orbital energies from a half-core hole calculation. The assumption is made that the excited electron does not participate in the decay of the core hole, and so the final-state energies are approximated by the ground state total energy with a single one-electron excitation using the ground state orbital energies. The unoccupied orbitals of the core excited states and the ground state are assumed to correspond by their energetic ordering. The transition dipoles for the absorption process are taken from the half core hole calculation and the ones for the emission process from the ground state calculation. All PESes are computed using the deMon2k code \cite{demon2k} making use of the newly implemented (by the author) interface of the code to the Atomic Simulation Environment \cite{Hjort-Larsen_2017}. The PBE functional \cite{PBE} is used with a TZVP basis set \cite{Godbout_1992} on the Carbon atoms, IGLO-II \cite{Kutzelnigg_book} on the hydrogens, and IGLO-III \cite{Kutzelnigg_book} on the oxygen that is core excited; the other oxygen is described with an effective core potential \cite{Igel-Mann_1988} along with a a triple-$\zeta$ valence [3s,3p,1d]  basis set. Although it can certainly be improved, the model will nevertheless enable a comparison between the quantum and semiclassical cases for this system. 

The parameters for the calculations are as follows. Transition dipole moments for all states are taken at the equilibrium geometry. A number of 10 intermediate states and 10 final states are included in the calculation. The HWHM lifetime broadening is set to 0.18 eV, the broadening of the incoming radiation is set to 0.9 eV and the instrumental broadening to 0.3 eV, all these values correspond to realistic experimental conditions \cite{Schreck_private_comm}. For the quantum case an additional Lorentzian broadening of 0.18 eV is employed in order to compare to the SCKH results. The polarization angle in Eq. (\ref{eq:pol_dep_theta}) is set to $\theta=0$.

Due to the relatively large width of the incoming radiation there is not as much need to sample the classical FC profile as for the HCl case. For this reason only four points were sampled each for the position and momentum which in total gives 16 trajectories. 

Since there is more than one intermediate state the question arises of the importance of doing the dynamics separately on each state, or if it can be done on a single more or less representative state. For applications, dynamics on a single state would mean an enormous simplification computationally and would make the cost of SCKH RIXS comparable to the cost of non-resonant SCKH XES. For this reason the KH spectrum is compared with two SCKH spectra, one with individual dynamics for each intermediate state and one with dynamics only on the first state. 
The intermediate state PESes are shown in the Supplementary Material  \cite{Sup_Mat}. 
As in the non-resonant case, the lowest PES is dissociative at the equilibrium OH position, around one \AA, but is bound by its hydrogen-bonded oxygen giving a minimum at around two \AA. The higher states are in general less dissociative but quite flat, and the general shape of the states is more of a particle in a box than a purely dissociative system like the previously treated HCl. It should be noted that there are some curve crossings 
for systems of this size
which means that one in general should take some kind of non-adiabaticity into account, however this is not within the scope of the present work. 

\begin{figure}[ptb]
\includegraphics[width=1.0\columnwidth, angle=0]{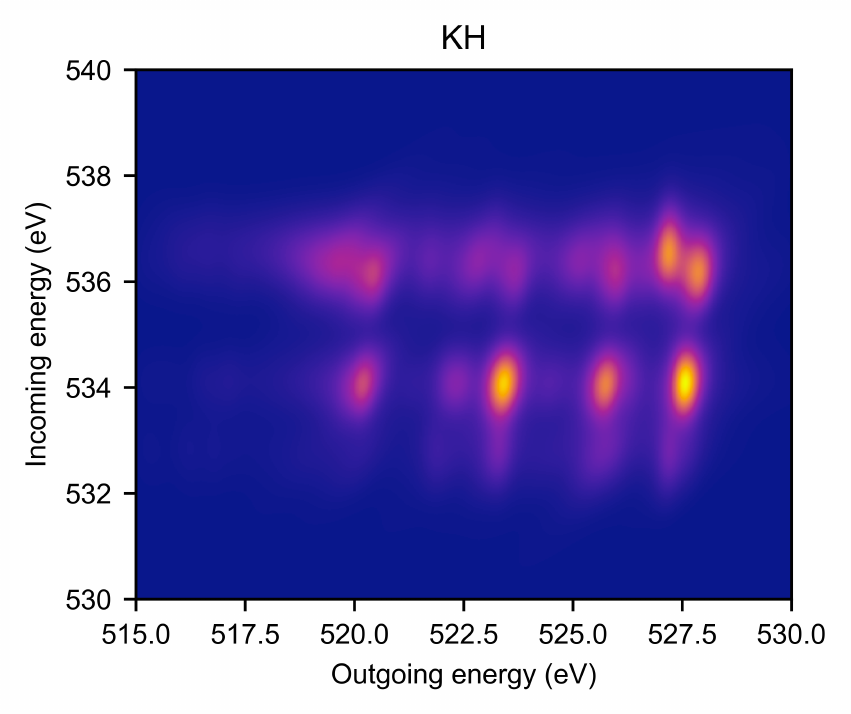}
\includegraphics[width=1.0\columnwidth, angle=0]{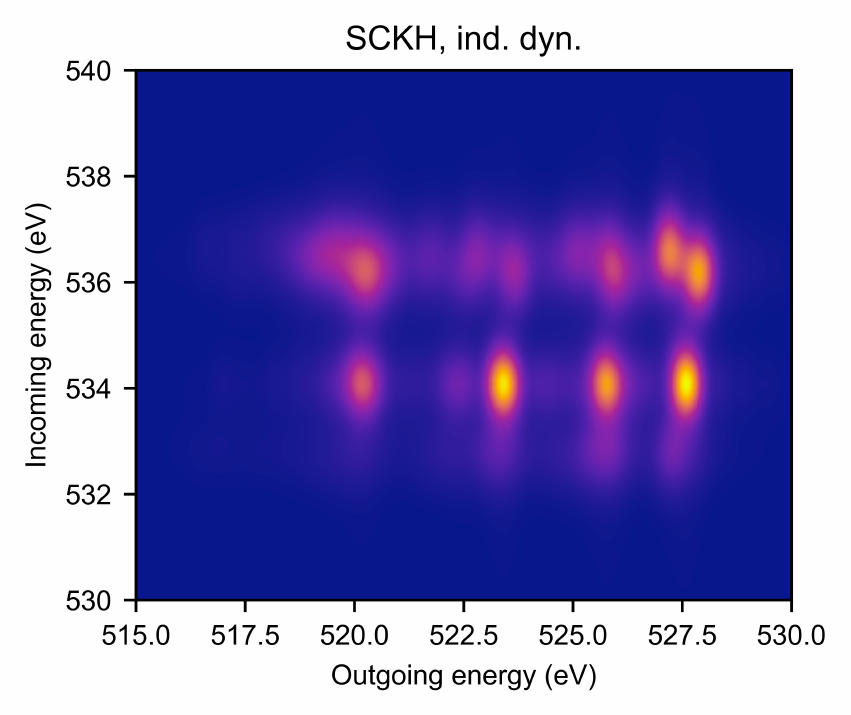}
\includegraphics[width=1.0\columnwidth, angle=0]{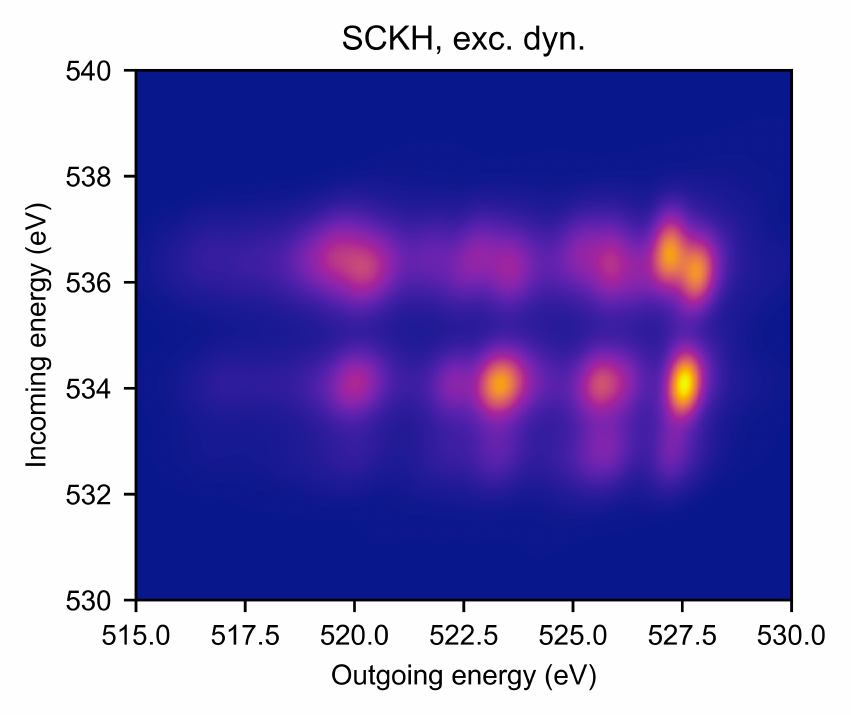}
 \caption{RIXS map for the methanol dimer. Upper frame: KH, middle frame: SCKH with individual dynamics, lower frame: SCKH with dynamics on the first core-excited state}
\label{fig:KH_res_methanol}
\end{figure}

RIXS maps for the three cases are shown in Fig. \ref{fig:KH_res_methanol} with, from the top down, KH, SCKH with individual dynamics on each intermediate state, and SCKH with dynamics on the first excited state only. The agreement of all methods is good, with the absorption and emission at more or less the right energy. The SCKH with individual dynamics reproduces the peak positions and intensities from the KH method well,
when using dynamics on the first excited state the intensities of the peaks 
become lower and 
slightly
smeared out. This should be expected when the PES where the dynamics is run on is more dissociative than the PES where the spectrum is computed and thus introduces too much dynamical effects. 

\begin{figure}[ptb]
\includegraphics[angle=0]{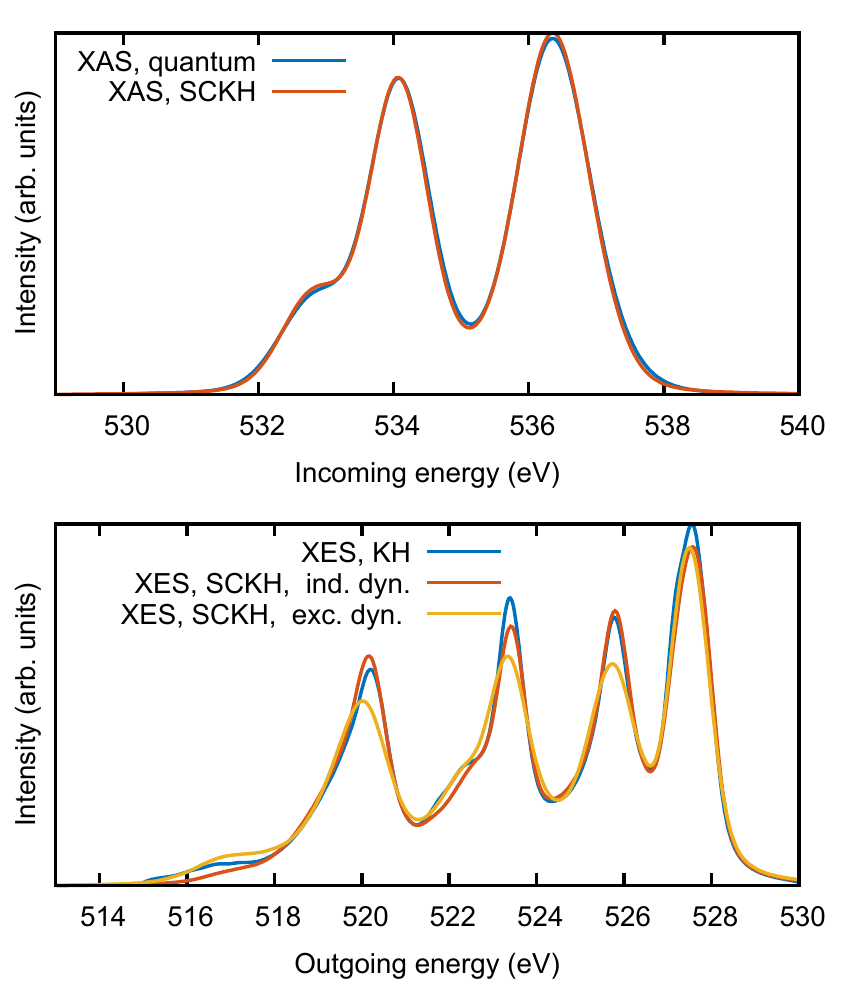}
 \caption{Averaged spectra for the methanol dimer. Upper frame: averaged spectrum over outgoing frequency. Blue: KH, red: SCKH. Lower frame: spectrum averaged over incoming frequency. Blue: KH, red: SCKH with individual dynamics, yellow: SCKH with dynamics on first excited state. The spectra are normalized.}
\label{fig:methanol_XAS_XES}
\end{figure}

In order to see the differences of the approaches more easily  $\langle \sigma(\omega) \rangle_{\omega'}$ and $\langle \sigma(\omega') \rangle_{\omega}$ are shown in Fig. \ref{fig:methanol_XAS_XES} . Looking first at the absorption in the upper frame, an almost perfect agreement between KH and SCKH is observed, which is not so surprising considering the success of the classical FC method for HCl and the harmonic test systems. The incoming broadening of HWHM of 0.9 eV clearly smears out the FC profile that would otherwise be seen in the quantum mechanical spectrum. For the XES (integrated over incoming frequency) the differences that are observed in the RIXS maps can be seen more clearly: with individual dynamics the features compared to KH 
have slightly wrong intensities.
For dynamics on the first excited state the 
lower peaks have too low and smeared-out intensities
Apparently the first state is too dissociative to describe the whole spectrum, and perhaps it could be wise to instead use some average state  for applications. It depends of course on which region of the emission spectrum that is of most interest. The agreement is, however, still very good. 

\begin{figure}[ptb]
\includegraphics[width=1.0\columnwidth, angle=0]{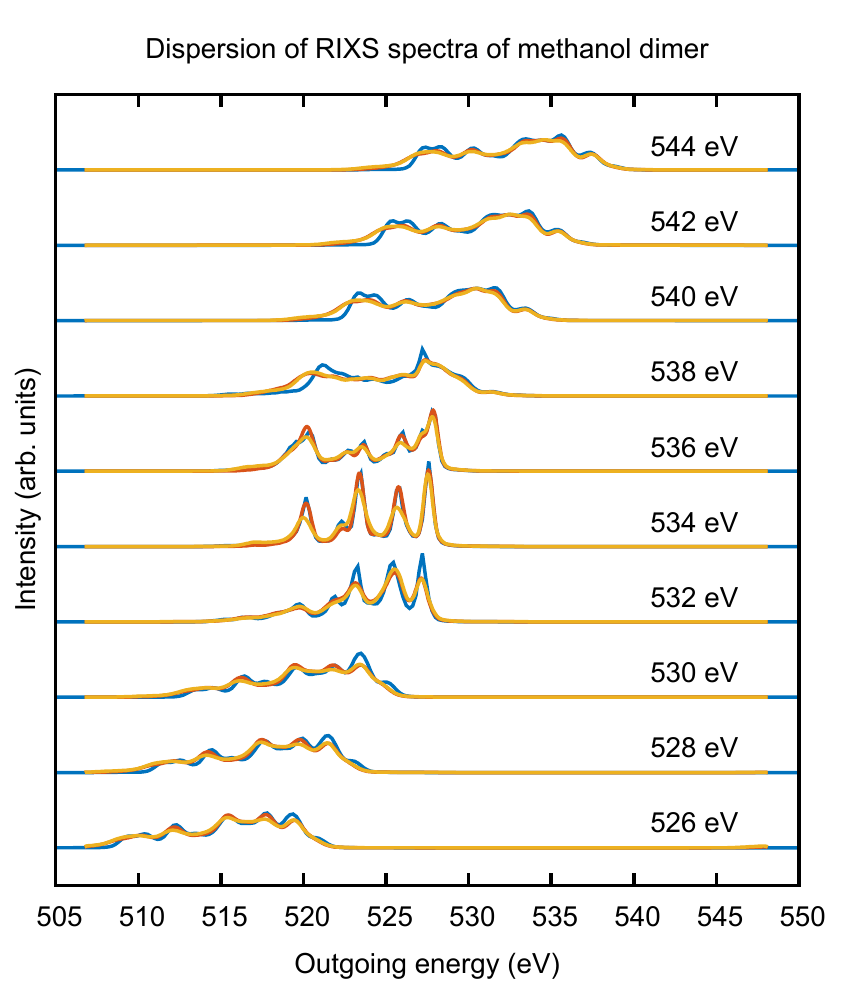}
 \caption{Normalized emission spectra for the methanol dimer. Blue: KH, red: SCKH with individual dynamics, yellow: SCKH with dynamics on first excited state. The absorption energy goes from 526 eV to 544 eV, bottom to top. }
\label{fig:dispersion_methanol}
\end{figure}

For the methanol dimer RIXS map there are no 
apparent
features that are linearly moving with the frequency, probably this is a combination of the larger lifetime broadening, the larger number of transitions, and the broader incoming radiation. Still, it would be interesting to see how the dispersion of the methanol dimer compares to that of HCl. In Fig. \ref{fig:dispersion_methanol} the normalized emission spectra for different incoming energies are shown, comparing the KH and SCKH (individual dynamics) results. Here the agreement is much better than for HCl, indeed is is almost as good as for the RIXS map and the averaged spectra. Apparently the many electronic transitions in combination with the larger broadening smears things out and gives less effects of unphysical features in the dispersion.  

It can thus be expected that for applications to liquids and other large systems that the dispersing features will be well described, and that the SCKH scheme will capture the important effects of vibrational interference, or nuclear dynamics on the spectra, and will be a relatively cheap alternative for such large systems. As the SCKH scheme can be used with any electronic structure method it is likely that a good, and cheap, description of the electric degrees of freedom will be key in accurately describe realistic systems. 

\section{Conclusions}

A new method is presented for computing resonant inelastic x-ray scattering (RIXS), that includes vibrational effects through classical dynamics on the core excited state. The method is an extension of the non-resonant semiclassical Kramers-Heisenberg formalism earlier developed in Ref.  \onlinecite{Ljungberg_2010}. To overcome difficulties with connecting the x-ray absorption and emission processes the classical Franck-Condon approximation is used for the absorption, which is seen to work very well for a series of harmonic test systems, as well as for more realistic test cases. 
The method gives qualitatively the right RIXS map for the case of L-shell RIXS of HCl, which is a difficult case since it  dissociates completely when core excited and has an extreme spectral feature corresponding to the dissociated geometry. For a more realistic case 
good agreement to the quantum results is obtained, provided that the dynamics is run on each intermediate state separately. If the dynamics is instead run on the first core excited state the description deteriorates somewhat, nevertheless the agreement is still good. Using a single state for the dynamics makes the computational cost of the method similar to the non-resonant case, that has already been successfully applied on liquid methanol \cite{Ljungberg_2017} and ethanol \cite{Takahashi_preprint_2017}. The dispersive part of the RIXS spectrum does not fully reproduce the quantum mechanical one, however, with many electronic states and large broadening the details are smeared out and a good agreement is found. In the case where vibrational effects are absent, the SCKH method reduces to the KH formula. 

The method presented in the present publication shows great  potential in modeling of liquids, such as alcohols and water solutions, and will enable also the resonant regime of the XES spectra to be explored, which have so far lacked good methods for modeling when vibrational effects are important. Further uses of the method could be for molecules adsorbed on surfaces, and in general complicated and disordered systems where a straightforward quantum mechanical approach of the vibrational degrees of freedom is not feasible. 

\section{Acknowledgements}

This work could not have been done without the constant encouragement and support of Lars G. M. Petterson, who
also provided valuable comments on the different versions of this manuscript.  
Iurii Zhovtobriuk and Elias Diesen are acknowledged for discussions and helping in the testing and running of the computer codes, and Peter Koval is thanked for providing computer time on his Linux cluster. 

%


\end{document}